\documentstyle[preprint,aps]{revtex}

\newcommand{\nc}{\newcommand}
\nc{\be}{\begin{equation}}
\nc{\ee}{\end{equation}}
\nc{\bea}{\begin{eqnarray}}
\nc{\eea}{\end{eqnarray}}
\nc{\beas}{\begin{eqnarray*}}
\nc{\eeas}{\end{eqnarray*}}
\nc{\noi}{\noindent}
\nc{\sD}{\not \! \! D}
\nc{\s}[1]{\not \! #1}
\nc{\non}{\nonumber}
\nc{\bb}{\bibitem}
\nc{\lf}{\left}
\nc{\ri}{\right}
\nc{\mb}[1]{\makebox[#1]{}}
\nc{\pa}{\partial}
\nc{\sA}{\not \! \! A}
\nc{\h}{\frac{1}{2}}
\nc{\ra}{\rightarrow}
\nc{\la}{\leftarrow}
\nc{\ep}{$e^+e^-\ra\pi^+\pi^-\;$}
\nc{\emuon}{$e^+e^-\ra\mu^+\mu^-\;$}
\nc{\epp}{$e^+e^-\ra\pi^+\pi^0\pi^-\;$}
\nc{\elec}{$e^+e^-\ra\gamma^*\ra e^+e^-\;$}
\def\mathunderaccent#1{\let\theaccent#1\mathpalette\putaccentunder}
\def\putaccentunder#1#2{\oalign{$#1#2$\crcr\hidewidth
\vbox to.2ex{\hbox{$#1\theaccent{}$}\vss}\hidewidth}}

\nc{\ti}{\mathunderaccent\widetilde}
\nc{\M}{{\cal M}}
\nc{\rw}{$\rho\!-\!\omega\;$}
\nc{\bold}[1]{\mbox{\boldmath $#1$}}
\nc{\lrpa}{\stackrel{\leftrightarrow}{\pa}}

\begin{document}

\tightenlines
\draft
\preprint{\vbox{\null \hfill  ADP-98-67/T334 \\
                                       \null \hfill UK/TP-98-16 \\
                                       \null\hfill hep-ph/9810422
                                       }}
\title{Hidden Symmetry and Georgi's ``Vector Limit"}
\author{{H.B. O'Connell}$^a$ and A.G.~Williams$^b$}
\address{$^a$Department of Physics and Astronomy, \\ 
	University of Kentucky,
        Lexington, KY 40506-0055 USA\\
        hoc@pa.uky.edu \\
$^b$Department of Physics and Mathematical
Physics \\
and Special Research Centre for the Subatomic Structure of Matter, \\
University of Adelaide 5005,   Australia  \\
  awilliam@physics.adelaide.edu.au}

\maketitle

\begin{abstract}

We study the generation of vector meson masses in 
the Hidden Local Symmetry (HLS)
model of low energy QCD.  
After demoting the HLS to a hidden global symmetry (HGS), the
$\rho$ is explicitly shown to be massless 
for any value of the HS parameter $a$ and
the chiral partner of the pion appears as the Goldstone boson associated
with the spontaneous breaking of the HGS,
closely resembling the predictions of Georgi's vector limit.  


{\it Keywords: 
Chiral symmetries,
Spontaneous symmetry breaking, 
Chiral Lagrangians, 
Vector-meson dominance.}

\end{abstract}

\pacs{PACS numbers: 
11.30.Rd,
11.30.Qc, 
12.39.Fe, 
12.40.Vv}

Recently there has been interest in the behaviour of $\rho$ mass 
under certain conditions
\cite{BR_MRS_LG_others,HarShi} and attention has turned to
Georgi's ``vector limit" in which the scalar and pseudoscalar fields
are on equal footing and the $\rho$ mass is presumed to
vanish \cite{georgi}.
Our purpose here is to examine Georgi's assumption that 
the  $\rho$ mass vanishes in the 
vector limit and discuss the more general issue of the role of the
$\rho$ as a ``dynamically generated" gauge boson \cite{bando}.

We shall begin with a very brief
outline of the Hidden Local Symmetry 
(HLS) model, introduced in Ref.~\cite{bando} and reviewed in
Ref.~\cite{BKYr}.
The HLS model
has the vector mesons as the gauge
bosons of a hidden local symmetry, as opposed to, say, 
a localised chiral symmetry (the so-called
``massive Yang-Mills model" see Ref.~\cite{BKYr}). 
The vector mesons acquire masses 
through the Higgs-Kibble (HK)
mechanism \cite{HK}. In this manner, the HLS
Lagrangian provides an accurate 
description of low energy QCD through its reproduction
of the phenomenologically successful vector meson
dominance (VMD) model (for a review see Ref.~\cite{review}). Let us
consider the chiral Lagrangian \cite{CCWZ},
\be
{\cal L_{\rm chiral}}=\frac{1}{4}{\rm Tr}
[\pa_\mu  F\pa^\mu F^{\dagger}],
\label{ccwz}
\ee
where 
\be
F(x)\equiv f_P U(x)\equiv 
f_P e^{2iP(x)/f_P },\,\,\,P(x)\equiv P^a(x)T^a
\ee
is the chiral field and the $SU(N_f)$ generator matrices
are normalised such that ${\rm Tr}[T^aT^b]=\delta^{ab}/2$.
This Lagrangian is invariant under 
\be
F\ra g_LFg_R^{\dagger},\,\,\,\,
g_{L,R}=e^{i\alpha_{L,R}^aT^a}
\ee
thus exhibiting 
the global symmetry $G=$SU$_L(N_f)\otimes$SU$_R(N_f)$, of 
massless QCD 
(for reviews see, e.g., Ref.~\cite{thomas}).
The chiral field can be expanded in terms of $P$
\be
F(x)=f_P e^{2iP(x)/f_P }=f_P +2iP(x)-2P^2(x)/f_P+\cdots\label{exp}
\ee
Substituting into Eq.(\ref{ccwz}) we see that the vacuum corresponds
to $P=0$, $F=f_P$. That is, $F$ has a non-zero vacuum expectation
value, $f_P$, which spontaneously breaks the 
symmetry of the vacuum
down to the subgroup ${\rm SU}_V(N_f)$ in which $g_L=g_R$. 
Spontaneous symmetry breaking is a necessarily non-perturbative
feature of QCD (for a discussion see, for example, Ref.~\cite{R}),
and in vector-like
gauge theories such as QCD, vector symmetries like the diagonal
subgroup, remain unbroken \cite{VW}.
The massless Goldstone bosons contained in $P$ are 
the pseudoscalars associated with this
spontaneous breaking of the coset space
$G/{\rm SU}_V(N_f)$ (also SU($N_f$)) and
correspond to perturbations
about the QCD vacuum.

A local symmetry, $H_{\rm local}$, can be added to this picture,
with group elements 
$h(x)=e^{ig\alpha^a(x)T^a}$. This is done
through the introduction of a scalar
``compensator field" $S$ transforming
like
\be
e^{iS(x)/f_S}\ra e^{iS^\prime(x)/f_S}=h(x)e^{iS(x)/f_S},\,\,\,\,
S(x)=S^a(x)T^a
\ee
where we note 
the interesting feature that 
a gauge transformation $h(x)=e^{-iS(x)/f_S}$
can completely remove the $S(x)$ field. As it can be gauged away, 
$S(x)$ is
unphysical.
Thus the ``hidden" local symmetry
can be included in Eq.~(\ref{ccwz}) by defining
\be
U(x)\equiv \xi^{\dagger}_L(x)\xi_R(x)\,,\,\,\,\,\,
\xi_{L,R}(x)=\exp[{iS(x)/{f_S}}]\exp[{\mp iP(x)/{f_P}}]
\ee
where
\be
\xi_{L,R}(x)\ra h(x)\xi_{L,R}(x)g^{\dagger}_{L,R}.
\ee

One now seeks to incorporate HLS into the low energy Lagrangian
in a non-trivial way, thereby introducing the lightest vector
meson states \cite{bando,BKYr}.
The procedure is to first rewrite ${\cal L_{\rm chiral}}$
explicitly in terms of the $\xi$ components  
\be
{\cal L_{\rm chiral}}=-\frac{f^2}{4}{\rm Tr}\left[
(\pa_\mu\xi_L\xi_L^{\dagger}-\pa_\mu\xi_R\xi_R^{\dagger})\right]^2.
\label{doon}
\ee
The Lagrangian can be gauged for both electromagnetism and the
hidden local symmetry by changing to covariant derivatives
\be
D_\mu\xi_{L,R}=\pa_\mu\xi_{L,R} -igV_\mu\xi_{L,R}+ie\xi_{L,R}A_\mu Q
\ee
where $Q$ is the charge matrix and
$V=V^aT^a$ transforms as $V_\mu\longrightarrow
h(x)V_\mu h^{\dagger}(x)+{i}h(x)
\pa_\mu h^{\dagger}(x)/{g}$. For full details of the matrices $P$ and
$V$ and the effects of flavour symmetry breaking 
for $N_f=3$ see, e.g., Refs.~\cite{before} and \cite{BO}.

{However, ${\cal L}_{A}\equiv{\cal L_{\rm chiral}}$ 
is independent of $V$, and a second Lagrangian, ${\cal L}_V$, is introduced}
\be
{\cal L}_{A}=-\frac{f_P^2}{4}{\rm Tr}\left[
(D_\mu\xi_L\xi_L^{\dagger}-D_\mu\xi_R\xi_R^{\dagger})\right]^2
,\,\,\,\,\,
{\cal L}_V=-\frac{f_P^2}{4}{\rm Tr}\left[
(D_\mu\xi_L\xi_L^{\dagger}+D_\mu\xi_R\xi_R^{\dagger})\right]^2
\label{mass1}
\ee
Both terms are invariant under $G$ and $H_{\rm local}$.
The full HLS Lagrangian is then given by  \cite{bando}
${\cal L}_{\rm HLS}={\cal L}_{A}+a{\cal L}_{V}$
introducing the HLS parameter $a$.

It is usual now to eliminate the scalars 
(which might be considered desirable as no chiral partner for the pion
has been observed) by
choosing the unitary 
gauge where,
\be \xi_L^{\dagger}(x)=\xi_R(x)\equiv\xi(x)=\exp[iP(x)/f_P].\label{P}\ee
However, 
the chiral transformation $g_{\rm ch}\in G$ 
will regenerate the scalar field through
\be
\xi(x)\ra\xi'(x)=\xi(x)g^{\dagger}_R
=g_L\xi(x)=\exp[iS'(P(x),g_{\rm ch})/f_S]\exp[iP'(x)/f_P],
\ee
and the system would no longer be in the unitary gauge. 
The local symmetry, though, allows for
Eq.~(\ref{P}) with the retention of 
chiral symmetry, $G$, under the combined
transformation \cite{BKYr}
\be
\xi(x)\ra\xi'(x)=h(P(x),g_{\rm ch})\xi(x)g_{\rm ch}^{\dagger},\hspace{1cm}
h(P(x),g_{\rm ch})=\exp[-iS'(P(x),g_{\rm ch})/f_S]
\ee
where the local
transformation, $h(P(x),g_{\rm ch})$,
``kills" the scalar field created by
the global transformation $g_{\rm ch}\in G$.
The transverse vector fields acquire
longitudinal components by ``eating" the scalar $S$ field through a
transformation of the form (to lowest order in Goldstone fields)
\be
V_\mu\ra V_\mu-\frac{1}{gf_S}\pa_\mu S'.\label{eat}
\ee
With this condition ${\cal L}_V$ produces the Lagrangian mass term,
$M_{\cal L}$, for the
vector mesons, 
\be
M_{\cal L}^2=af_P^2g^2.\label{Lm}
\ee  
The interaction terms 
of the $\rho$, photon and pions are given by
\be
{\cal L}^{\rm int}=
i[({a}/{2})g\rho_\mu+(1-a/2)eA_\mu](\pi^-\pa^\mu\pi^+-\pi^+\pa^\mu\pi^-)
-eaf_P^2g^2\rho_\mu A^\mu.
\ee
The choice $a=2$ \cite{bando} gives the usual formulation of VMD,
fixing $g=g_{\rho\pi\pi}$ and removing the direct photon-pion
contact term (the photon can only couple to the pseudoscalars through
the vector mesons).
Eq.~(\ref{Lm}) then reproduces the KSRF relation \cite{KSRF}.

Let us now discuss exactly how the vectors acquire mass, and the
role of the scalar field in this in order to 
understand Georgi's massless vector limit and associated claim of
a vanishing coupling constant through
\be
m_\rho/f_P\ra0\Rightarrow g\ra0
\ee
as might be inferred from Eq.~(\ref{Lm}).
We shall continue to work in the HLS model and
restrict our 
specific attention to the neutral $\rho$.
The relevant Lagrangian term
is given by
\be
{\cal L}_{SV}=af_P^2{\rm Tr}[gV_\mu-\pa_\mu S/f_S]^2.\label{SV}
\ee
The gauge symmetry, $H_{\rm local}$, is spontaneously broken
by the vacuum expectation value of the compensator field
$e^{iS(x)/f_S}$, and the field $S$ is the 
``would be" Goldstone boson associated
with this.
The gauge fixing in Eq.~(\ref{eat}) removes the $S$
fields and reduces Eq.~(\ref{SV})
to simply the Lagrangian mass term of the
vector field.  The vector limit cannot be realised, because the
scalar field is unphysical, and so cannot appear as the chiral
partner of the pion.

However, as Georgi points out \cite{georgi}, there is no physics in the
hidden local symmetry.  Indeed, while global symmetries are associated with
observable currents, gauge symmetries merely represent a redundancy in the
description of a physical system \cite{S}.  Let us now consider demoting
the HLS $H_{\rm local}$ to a hidden {\em global} symmetry (HGS), $H_{\rm
global}$, but keep the vector fields which transform now as $V_\mu\ra
hV_\mu h^\dagger$ ($h\in H_{\rm global}$).  Naturally, ${\cal L}_A$ is
unaffected by this:  we still have $G$ broken to SU(3)$_V$ by the vacuum
expectation value of the $F(x)$ fields.  Let us concentrate on ${\cal
L}_V$.  In this case the vacuum expectation value of the field $e^{iS(x)}$,
transforming as $e^{iS(x)}\ra he^{iS(x)}$, spontaneously breaks the $H_{\rm
global}$ symmetry (the $e^{iP(x)}$ are singlets under $H_{\rm global}$ just
as $e^{iS(x)}$ is a singlet under $G$) and the $S(x)$ are now legitimate
Goldstone bosons.

We shall now consider the effect of this on the vector meson masses,
in particular that of the neutral $\rho$, which is of interest
as Georgi describes a vanishing $\rho$ mass in his vector
limit \cite{georgi}. 
Introducing the chiral partner of the pion, 
$\sigma$ (an isospin triplet), which is an element of $S$ the way $\pi$
is an element of $P$,
Eq.~(\ref{SV}) gives
\be
{\cal L}_{\rho\sigma}=\frac{1}{2}af_P^2g^2\rho^2
+\frac{1}{2}\frac{af_P^2}{f_S^2}
(\pa\sigma)^2-g\frac{af^2_P}{f_S}\rho_\mu\pa^\mu\sigma.\label{RS}
\ee
The first term on the RHS of Eq.~(\ref{RS}) is simply the Lagrangian 
mass term of the $\rho$ generated by the
vacuum expectation value of the field $e^{iS(x)}$.
The second term is the kinetic term for the $\sigma$ meson.
The third term couples
the massless $\sigma$ to
the $\rho$. We now see that
Eq.~(\ref{RS})
suggests a slightly more general limit than
Georgi's $a=1$. So that the $\sigma$ appears as a physical particle with
a normalised kinetic energy term we have,
\be
f_S^2={a}f_P^2\label{f_S}.
\ee
We shall refer to this
as the BKY limit (following Sect.~6 of Ref.~\cite{BKYr}).
However, $f_S$ which has appeared here so far is merely a 
parameter. Let us relate it to the physical scalar decay constant
of the $\sigma$,
which defines Georgi's vector limit through \cite{georgi}
\be
\langle0|A^\mu|\pi\rangle=if_\pi p^\mu ,\,\,\,\,\,
\langle0|V^\mu|\sigma\rangle=if_\sigma p^\mu ,\,\,\,\,\,
f_\pi=f_\sigma,
\ee
The extraction of the
pseudoscalar decay constants from the HLS Lagrangian is discussed in
Section IV.A of Ref.~\cite{BO}. Following this
we find,
\be
f_\sigma=\frac{af_P^2}{f_S}=\frac{af_P^2}{f_S^2}f_S,\,\,\,\,
f_\pi=f_P
\ee
and we find the BKY limit (Eq.~(\ref{f_S})) yields $f_\sigma=f_S$.
Setting $a=1$, would then give Georgi's $f_\sigma=f_P$ \cite{georgi},
though this value for $a$ seems disfavoured by data \cite{BEMOSW}.

We are now in a position to determine the {\em physical} mass of the
$\rho$. 
Firstly, let us state that we shall ignore the effects of 
pseudoscalar loops in the
following analysis. These terms contribute to the
vector pole through the polarisation function, $\Pi(s)$, 
which we define as being generated solely through meson loops.
The physical pole, $s=p_V$ is then given by $p_V-m^2-\Pi(p_V)=0$.
The polarisation functions develop an imaginary part above the
pseudoscalar production threshold and move the pole of the
vector meson propagator from the real axis 
to the complex plane, thus generating
the physically important meson widths \cite{BOW,GO}. However, 
they have little effect on the real part of the mass term: 
numerical studies find the real part of the $\rho$ mass to
be typically altered only by a few percent, and even less for the
$\omega$ \cite{loops}. Furthermore, as we are interested in examining
Georgi's statement that the vector meson masses vanish, our interest
lies in the vector propagators around $s=0$. 
In a model such as the HLS model, the vectors couple to conserved
currents, as can be seen form the interaction Lagrangian given, for
example, in the Appendix of Ref.~\cite{BO}. The vector currents take
the form $J_\mu=a\pa_\mu b-b\pa_\mu a$, where $a$ and $b$ are 
pseudoscalars and thus $\pa_\mu J^\mu=0$ from the pseudoscalar equations
of motion $\pa^2a=\pa^2b=0$. As  $\Pi(s)$ is generated from couplings
to conserved currents
we have $\Pi(s=0)=0$ \cite{node}. 

Therefore, the tree level is entirely adequate
for our purposes.
The physical mass is then
determined by two things.
The first, naturally, is $M_{\cal L}$, the second is
the dressing of the
$\rho$ propagator by the $\sigma$
(a similar effect is discussed for the weak bosons
by Farhi and Susskind \cite{FS} and Peyranere \cite{P}
and for the $\pi-a_1$ system by Kaloshin \cite{K}).
We have 
\be
(D_{\mu\nu}^\rho)^{-1}=(M_{\cal L}^2-s+M_{\sigma\rho}^2)g_{\mu\nu}
+q_\mu q_\nu.\label{rs}
\ee
The contribution to the $\rho$ dressing from the $\sigma$ is given by
\be\label{qq}
iM_{\rho\sigma}^2=\left(-ag\frac{f_P^2}{f_S}q_\mu\right)\frac{i}{q^2}
\left(ag\frac{f_P^2}{f_S}q^\mu\right)\Rightarrow
M_{\rho\sigma}^2=-a^2g^2\frac{f_P^4}{f_S^2}.
\ee
The mass contribution $M_{\rho\sigma}$ is seen to be {\em imaginary},
due to the relative minus sign between the $\sigma$
momentum, $q_\mu$, going {\em in} to each $\rho\sigma$ 
vertex in Eq.~(\ref{qq}). This is much like what
Sakurai discovered for the vector meson contribution to the vacuum
polarisation of the photon \cite{Sak}.  Sakurai realised that this could
be cancelled by a Lagrangian mass term for the photon
(see also Ref.~\cite{BOW}). Such a term
is provided automatically in the HLS model, $M_{\cal L}$.
The physical mass for the $\rho$,
\be
(m^{\rm phys}_\rho)^2=M_{\cal L}^2+M^2_{\rho\sigma}=ag^2f_P^2-ag^2f_P^2
\frac{af_P^2}{f_S^2},\label{vanish}
\ee 
which indeed vanishes in the BKY limit (Eq.(\ref{f_S}))
 {\em for any value of $a$}. 
One sees the vanishing $\rho$ mass
does not require $g\ra0$
\cite{georgi,BR2}.
It comes automatically after one demotes HLS to HGS, which gives
rise to a chiral partner for the pion \cite{georgi,SK}.
Indeed, 
$g\ra0$ is merely the extreme case of our demotion of HLS to HGS, namely
removal of the symmetry altogether ($h=I$).
The vanishing of the $\rho$ mass
in the BKY limit
arises from the cancellation 
between the $\sigma$ meson
contribution to the $\rho$ vacuum polarisation Lagrangian mass term
for the $\rho$. 
Therefore previous calculations where gauge freedom can be used to remove the
$\rho-\sigma$ interaction \cite{HY}
will miss this behaviour and describe a massive
$\rho$ \cite{HarShi,RCW}. Our choice of a 
hidden {\em global} symmetry, makes
the $\rho-\sigma$ interaction unavoidable because the $\sigma$ cannot
be gauged away.

In conclusion, we note that the point of this work has been
to examine Georgi's ``vector limit"  within the context of
the hidden local symmetry  model of the low energy hadronic
sector.
By demoting the hidden local symmetry
to a hidden {\em global} symmetry  we have reproduced
the key characteristics of the vector limit.
We have shown that a chiral partner of the pion,
which we refer to as the $\sigma$,  appears as the 
Goldstone boson associated with the spontaneous breaking of the
hidden global symmetry. Due to the contribution of the scalar $\sigma$
to the vacuum polarisation of the $\rho$, the $\rho$ is  
massless, as assumed by Georgi in the vector limit, but {\em without}
the $\rho$ coupling, $g$, also vanishing. In our hidden global
symmetry model
we find $f_S=\sqrt{a}f_P$ due to the normalisation of the
scalar kinetic term, and Georgi's
vector limit constraint $a=1$ serves only to set $f_S=f_P$.

\begin{center}
{\bf Acknowledgements}
\end{center}
We would like to thank T.D.~Cohen,
K.F.~Liu, J.~McCarthy, J.~Sloan and W.~Wilcox
for helpful correspondence and discussions.
This work is supported by the
US Department of Energy under grant DE--FG02--96ER40989 (HOC)
and the Australian Research Council (AGW).

\end{document}